\documentclass[11pt]{article}
\usepackage{graphicx,amsmath,amsfonts,amssymb,dcolumn,times}

\begin{document}

\title{\textbf{The role of gauge symmetry in spintronics}}
\author{\textbf{R.~F.~Sobreiro$^1$}\thanks{sobreiro@if.uff.br}\ , \textbf{V.~J.~Vasquez Otoya$^2$}\thanks{victor.vasquez@ifsudestemg.edu.br}\\\\
\textit{{\small $^1$UFF $-$ Universidade Federal Fluminense,}}\\
\textit{{\small Instituto de F\'{\i}sica, Campus da Praia Vermelha,}}\\
\textit{{\small Avenida General Milton Tavares de Souza s/n, 24210-346,}}\\
\textit{{\small Niter\'oi, RJ, Brasil.}}\and
\textit{{\small $^2$IFSEMG $-$ Instituto Federal de Educa\c{c}\~ao, Ci\^encia e Tecnologia do Sudeste de Minas Gerais,}} \\
\textit{{\small Rua Bernardo Mascarenhas 1283, 36080-001,}}\\
\textit{{\small Juiz de Fora, MG, Brasil}}}
\date{}
\maketitle

\begin{abstract}
In this work we employ a field theoretical approach to explain the nature of the non-conserved spin current in spintronics. In particular, we consider the usual U(1) gauge theory for the electromagnetism at classical level in order to obtain the broken continuity equation involving the spin current and spin-transfer torque. Inspired in the recent work of A. Vernes, B. L. Gyorffy and P. Weinberger where they obtain such equation in terms of relativistic quantum mechanics, we formalize their result in terms of the well known currents of field theory such as the Bargmann-Wigner current and the chiral current. Thus, an interpretation of spintronics is provided in terms of Noether currents (conserved or not) and symmetries of the electromagnetism. In fact, the main result of the present work is that the non-conservation of the spin current is associated to the gauge invariance of physical observables where the breaking term is proportional to the chiral current. Moreover, we generalize their result by including the electromagnetic field as a dynamical field instead of an external one.
\end{abstract}

\section{Introduction}

Spintronics or spin-based electronics is part of the study of transport phenomena in condensed matter Physics, it explores the spin properties of electrons and its  understanding  plays an important role in the development of new electronic devices. Interesting effects as giant magnetoresistance and spin-Hall are extensively explored  in order to produce some technological applications \cite{Baibich:1988zz,Binasch:1989zz,Sinova}. Nevertheless, the non-conservation of spin current continue being a problem \cite{Qing,Vernes:2007}, at least from the theoretical point of view. The purpose of this work is not to solve that problem but to understand the origin of this non-conservation law under the theoretical framework of gauge field theories.

It is known that some couplings in condensed matter can be understood as non-relativistic effects of a relativistic theory, for instance the non-relativistic limit of quantum electrodynamics predicts the existence of spin-orbital coupling and the exact value for the gyromagnetic factor which is a classic example \cite{Itzykson:1980rh}. In \cite{Vernes:2007} the same technique was employed to obtain the broken continuity equation for spin currents. In fact, this equation is derived from the non-relativistic limit of the time evolution of the Bargmann-Wigner operator \cite{Bargmann:1948ck}, resulting in
\begin{equation}
\frac{d\overrightarrow{s}}{dt}+\partial_i\overrightarrow{j}_i=\frac{e}{m}\overrightarrow{s}\times\overrightarrow{B}\;,\label{paulidirac}
\end{equation}
where $\overrightarrow{s}=\phi^\dagger\overrightarrow{\sigma}\phi$ is the spin density, $\overrightarrow{j}_i$ is the spin current, $\phi$ is the non-relativistic electron wave function and the \emph{rhs} is the usual microscopic Landau-Lifshitz torque. The second term in the \emph{lhs} of \eqref{paulidirac} is the spin-transfer torque. The very same equation \eqref{paulidirac} can also be obtained directly from the study of time evolution of $\overrightarrow{s}$ if one considers the Pauli equation for the electron \cite{Stiles}. It must be remarked that equation \eqref{paulidirac} can also be obtained from the gauge current of the $SU(2)$ gauge theory (flavor current in weak interactions) by decomposing its continuity equation and rearranging the fields \cite{Dartora:2008ccc, Dartora:2010zz}. Thus, spin current flow can actually be casted into a continuity equation even though standard electromagnetism based on the $U(1)$ gauge symmetry is substituted by its next non-Abelian extension.

Roughly speaking, an electron is fundamentally characterized by its mass, charge and spin. These properties could be better understood from the point of view of symmetries in field theory \cite{Itzykson:1980rh}, in which the Poincar\'e group determines all  information about the mass and spin of the particles; in fact the eigenvalues of Casimir operators related to the Poincar\'e group provides the mass and spin values of particles, which are irreducible representations of Poincar\'e group \cite{Itzykson:1980rh}. The mass is computed from the translational sector of the group while the spin is related to the subgroup that leaves the classical linear momentum invariant, the so called little group \cite{Itzykson:1980rh,Bargmann:1948ck}. On the other hand, the charges of these particles are related to the internal symmetries; the interaction process between matter and fields is realized when these internal symmetries are localized, giving birth to a gauge theory.

In the present paper we consider the Maxwell-Dirac action and the currents associated with the gauge symmetry, chiral asymmetry and little group symmetry. The latter is commonly known as Bargmann-Wigner current whose generators are precisely those associated with the values of spin in the same way that the generators of the subgroup of translations are associated to the mass value of particles. Although conserved, the Bargmann-Wigner current is not gauge invariant and thus hard to be associated with physical observables. Thus, by evoking the gauge principle for electrodynamics we propose to restore the gauge invariance of the Bargmann-Wigner by substituting the ordinary derivative by the covariant one and thus associate it with a physical observable. The price that is payed is that the current is no longer conserved where the breaking term is identified with spin torque. Thus, those effects are regarded as natural consequences of this non-conservation law. Moreover, since the electromagnetic field is dynamical we also obtain a similar equation for the photon. In that case it is shown that the photon spin current can be used to measure how the angle between electric and magnetic fields differs from $\pi/2$. Also, if this current is non-zero it can be associate associate with a kind of magnetic monopole force.

The paper is organized as follows: In Section 1 we define the starting action, the fields and symmetries of electrodynamics (All this content are well know from text books, however it is necessary ti fix the notation and conventions). Section 3 is devoted to the restoration of gauge invariance of the Bargmann-Wigner current and the corresponding broken spin continuity equations. In Section 4 we provide the decomposition of these equations in space and time sector. Also in this section, the interpretation of such equations are performed. Finally, the conclusions are displayed in section 5.

\section{Maxwell-Dirac action and its symmetries}

We begin with the standard electromagnetic action, in four Minkowiskian dimensions,
\begin{equation}
S=S_o+S_{int}+S_M\;,\label{action}
\end{equation}
where $S_o$ is the Dirac fermionic action, $S_{int}$ the interacting term and $S_M$ the Mawell action, namely
\begin{eqnarray}
S_o&=&\int{d^4x}\;\overline{\psi}\left(i\gamma^\mu\partial_\mu-m\right)\psi\;,\nonumber\\
S_{int}&=&-e\int{d^4x}\;\overline{\psi}\gamma^\mu\psi
A_\mu\;,\nonumber\\
S_M&=&-\frac{1}{4}\int{d^4x}\;F_{\mu\nu}F^{\mu\nu}\;.\label{actions}
\end{eqnarray}
The field strength is defined as $F_{\mu\nu}=\partial_\mu A_\nu-\partial_\nu A_\mu$, where $A_\mu$ is the electromagnetic potential. The quantity $e$ stands for the electromagnetic coupling constant. The field $\psi$ is a spinor field describing for electron excitations
and $\overline{\psi}$ its adjoint, $\overline{\psi}=\psi^\dagger\gamma^0$. The Clifford algebra $\left\{\gamma^\mu,\gamma^\nu\right\}=2\eta^{\mu\nu}$ allows the usage of Dirac representation for the $\gamma$-matrices and the metric tensor is defined with negative signature, $\eta=\mathrm{diag}(+1,-1,-1,-1)$. Useful extra quantities are $\gamma^5=\gamma_5=i\gamma^0\gamma^1\gamma^2\gamma^3$ and $\sigma^{\mu\nu}=\frac{i}{2}\left[\gamma^\mu,\gamma^\nu\right]\label{sigma}$.

From \eqref{action}, one can extract the spinor field equations
\begin{eqnarray}
\left(i\gamma^\mu D_\mu-m\right)\psi&=&0\;,\nonumber\\
\overline{\psi}\left(i\gamma^\mu\overleftarrow{D}_\mu^\dagger+m\right)&=&0\;,\label{Eqf}
\end{eqnarray}
where the covariant derivative is a diagonal matrix $D_\mu=\partial_\mu+ieA_\mu$.  For the electromagnetic field, the equations are
\begin{equation}
\partial_\nu F^{\nu\mu}=e\overline{\psi}\gamma^\mu\psi\;,\label{Eqb}
\end{equation}
recognized as the inhomogeneous Maxwell equations. The homogeneous Maxwell equations, $\partial_\nu
\widetilde{F}^{\nu\mu}=0$, are obtained from the topological properties of the theory, where the dual field strength is defined as $\widetilde{F}^{\mu\nu}=\frac{1}{2}\epsilon^{\mu\nu\alpha\beta}F_{\alpha\beta}$.

The $U(1)$ local symmetry is characterized by the set of transformations
\begin{eqnarray}
\delta_g\psi&=&-ie\alpha\psi\;,\nonumber\\
\delta_g\overline{\psi}&=&ie\alpha\overline{\psi}\;,\nonumber\\
\delta_gA_\mu&=&\partial_\mu\alpha\;,\label{U1}
\end{eqnarray}
where $\alpha$ is a spacetime dependent parameter. The action \eqref{action} is invariant under gauge transformations and the associated off-shell conserved current is\footnote{We are using the following definition for the Noether current
\begin{equation}
J^\mu=\frac{\partial\mathcal{L}}{\partial(\partial_\mu \phi^A)}\delta\phi^A-\left(\frac{\partial\mathcal{L}}{\partial(\partial_\mu \phi^A)}\partial_\nu\phi^A-\delta^\mu_\nu\mathcal{L}\right)\delta x_\nu\;,\nonumber
\end{equation}
where $\mathcal{L}$ is the Lagrangian and $A$ a general collective index characterizing the indices of the fields as well as the sum over all fields.}
\begin{equation}
j_f^\mu=e\overline{\psi}\gamma^\mu\psi\;,\label{charge}
\end{equation}
It follows from Noether's theorem that $\partial_\mu j^\mu_f=0$, expressing the electric charge conservation.

Another well known fact in electrodynamics is the presence of the
chiral symmetry for massless fermions. The Chiral transformations
are defined as
\begin{eqnarray}
\delta_c\psi&=&-i\alpha\gamma^5\psi\;,\nonumber\\
\delta_c\overline{\psi}&=&-i\alpha\overline{\psi}\gamma^5\;,\nonumber\\
\delta_cA_\mu&=&0\;,
\end{eqnarray}
where $\alpha$ is now a constant parameter. The chiral non conserved current is easily computed as
\begin{equation}
S^\mu=\overline{\psi}\gamma^\mu\gamma^5\psi\;,\label{chiral}
\end{equation}
which leads to the broken continuity equation $\partial_\mu S^\mu=2im\overline{\psi}\gamma^5\psi$.

Finally, the third set of transformations relevant to this work originates from the little group sector of the Lorentz group, $H\subset SL(2,\mathbb{C})$. The generators of the Poincar\'e group are denoted by $P_\mu=i\partial_\mu$ for translations and $J_{\mu\nu}=L_{\mu\nu}+I_{\mu\nu}$ for the Lorentz sector. Here, $L_{\mu\nu}$ is taken as the angular momentum
part, $L_{\mu\nu}=i\left(x_\mu\partial_\nu-x_\nu\partial_\mu\right)$, while $I_{\mu\nu}$ is associated to the internal angular momentum, \emph{i.e.}, the spin. The Poincar\'e algebra is:
\begin{eqnarray}
\left[P_\mu,P_\nu\right]&=&0\;,\nonumber\\
\left[J_{\mu\nu},J_{\alpha\beta}\right]&=&\eta_{\mu\alpha}J_{\nu\beta}-\eta_{\mu\beta}J_{\nu\alpha}-\eta_{\nu\alpha}J_{\mu\beta}+ \eta_{\nu\beta}J_{\mu\alpha}\;,\nonumber\\
\left[J_{\mu\nu},P_\alpha\right]&=&i\eta_{\alpha\nu}P_\mu-\eta_{\alpha\mu}P_\nu\;.\label{sl2c}
\end{eqnarray}
The little group can be understood as a Lorentz transformation that maintain invariant the linear momentum, \emph{i.e.}
\begin{equation}
W^\mu=-\frac{1}{2}\epsilon^{\mu\nu\alpha\beta}J_{\nu\alpha}P_\beta=-\frac{1}{2}\epsilon^{\mu\nu\alpha\beta}I_{\nu\alpha}P_\beta\;.
\end{equation}
Thus
\begin{eqnarray}
\left[W^\mu,W^\nu\right]&=&0\;,\nonumber\\
\left[J_{\mu\nu},W^\alpha\right]&=&i\left(\delta^\alpha_\nu\eta_{\mu\beta}-\delta^\alpha_\mu\eta_{\nu\beta}\right)W^\beta\;,\nonumber\\
\left[W^\mu,P_\alpha\right]&=&0\;,\label{little}
\end{eqnarray}
emphasizing the subgroup character of the little group as well as the fact that $H$ is a stability subgroup of the Poincar\'e group.

For fermions, it is easy to find \cite{Itzykson:1980rh}
$I_{\mu\nu}=\sigma_{\mu\nu}/2$, providing
\begin{equation}
W^\mu_f=-\frac{1}{4}\epsilon^{\mu\nu\alpha\beta}\sigma_{\nu\alpha}P_\beta=\frac{i}{2}\gamma^5\sigma^{\mu\nu}P_\nu\;,\label{Pauli-Lubanski}
\end{equation}
recognized as the Pauli-Lubanski vector where the index $f$ denotes its fermionic character. The little group transformations are
then
\begin{eqnarray}
\delta_l\psi&=&-i\omega_\mu W^\mu_f\psi\;,\nonumber\\
\delta_l\overline{\psi}&=&-i\overline{\psi}\overleftarrow{W}^\mu_f\omega_\mu\;,\label{little0}
\end{eqnarray}
with $\omega_\mu$ a set of constant real parameters. The related Noether
current is a second rank tensor field, the so called Bargmann-Wigner
tensor
\begin{equation}
T_f^{\mu\nu}=\overline{\psi}\gamma^\mu W^\nu_f\psi\;.
\end{equation}

For the electromagnetic field, the Pauli-Lubanski vector reads\footnote{It follows from the spin part of the Lorentz generator for a vector field,
$\sigma^{\mu\nu\alpha\beta}=(\eta^{\mu\alpha}\eta^{\nu\beta}-\eta^{\mu\beta}\eta^{\nu\alpha})/2$.}
\begin{equation}
W_b^{\mu\nu\alpha}=-\frac{1}{2}\epsilon^{\mu\nu\alpha\beta}P_\beta\;,
\end{equation}
where the index $b$ characterizes its bosonic behavior. The little group transformation for $A_\mu$,
\begin{equation}
\delta_lA_\mu=i\omega^\nu W_{b\;\mu\nu\alpha}A^\alpha\;.\label{little1}
\end{equation}
Thus, for the vector field the corresponding Bargmann-Wigner current is
\begin{equation}
T_b^{\mu\nu}=\frac{1}{2}F^{\mu\alpha}\widetilde{F}_\alpha^{\phantom{\alpha}\nu}\;,
\end{equation}
and the full Bargmann-Wigner conserved current is then
\begin{equation}
T^{\mu\nu}=T_f^{\mu\nu}+T_b^{\mu\nu}\;\;\bigg|\;\;\partial_\mu T^{\mu\nu}=0\;.\label{barg}
\end{equation}

\section{Restoring the gauge invariance of Bargmann-Wigner tensor}

It turns out that, in contrast to gauge and chiral currents, $T^{\mu\nu}$ is not a gauge invariant quantity,
\begin{equation}
\delta_gT^{\mu\nu}=\delta_gT_f^{\mu\nu}=-ie\overline{\psi}\gamma^\mu\left(W^\nu\alpha\right)\psi\;,
\end{equation}
and thus, from the gauge principle, it cannot be associated to a physical observable. The gauge nature of electrodynamics automatically forbids $T^{\mu\nu}$ to be a physical quantity. Moreover, it is evident that it is only the fermionic sector $T^{\mu\nu}_f$ that breaks gauge symmetry. To circumvent this problem we generalize $T_f^{\mu\nu}$ to its simplest gauge invariant extension by replacing the ordinary derivative by the covariant one, \emph{i.e.}, the Pauli-Lubanski vector is replaced by
\begin{equation}
\mathcal{W}^\mu_f=-\frac{1}{2}\gamma^5\sigma^{\mu\nu}D_\nu\;,
\end{equation}
implying that the gauge invariant fermionic Bargmann-Wigner current is now
\begin{equation}
\mathcal{T}^{\mu\nu}_f=\frac{1}{2}\overline{\psi}\gamma^5\gamma^\mu\sigma^{\nu\alpha}D_\alpha\psi\;,\label{little1}
\end{equation}
Thus, since the electromagnetic sector is already gauge invariant, the full gauge invariant Bargmann-Wigner current is
\begin{equation}
\mathcal{T}^{\mu\nu}=T^{\mu\nu}+\frac{ie}{2}\overline{\psi}\gamma^5\gamma^\mu\sigma^{\nu\alpha}\psi
A_\alpha\;,\label{little2}
\end{equation}
and now $\delta_g\mathcal{T}^{\mu\nu}=0$.

If in one hand we have the gauge invariance, in the other hand
$\mathcal{T}^{\mu\nu}$ is not conserved anymore. However, it is
indeed an observable of the theory, just like $S^\mu$ which is not
conserved but it is gauge invariant. In fact, it can be straightforward shown that
\begin{equation}
\partial_\nu\mathcal{T}_f^{\nu\mu}=\frac{e}{2}S_\nu
F^{\nu\mu}\;,\label{div1}
\end{equation}
where the field equations \eqref{Eqf} were used. For the bosonic sector it is easy to show that,
\begin{equation}
\partial_\nu T^{\nu\mu}_b=\frac{1}{2}\widetilde{j}_f^{\mu\nu\alpha}F_{\nu\alpha}\;,\label{div2}
\end{equation}
where \eqref{Eqb} was used and $\widetilde{j}^{\mu\nu\alpha}_f$ is the dual $U(1)$ current, $\widetilde{j}^{\mu\nu\alpha}_f=\epsilon^{\mu\nu\alpha\beta}j_\beta/3!$.

A few comments are in order: First, equations \eqref{div1} and \eqref{div2} expresses the non conservation of the gauge invariant
Bargmann-Wigner currents, and hold separately, since they are obtained independently of the continuity equation \eqref{barg}. Second, equation \eqref{div1} is in fact the covariant version of the result obtained in \cite{Vernes:2007}, which associates the Bargmann-Wigner current with spin currents and spin-torque transfer effects when the electromagnetic field is strictly external. Third, equation \eqref{barg} is an extra equation, expressing the little group symmetry, that always holds independently of \eqref{div1} and \eqref{div2}. 

\section{Spacetime decomposition}

In order to provide a more intuitive understanding of \eqref{div1} and \eqref{div2} it is useful to decompose these equations into space and time sectors. We start by considering fermionic equation \eqref{div1}. Following \cite{Vernes:2007}, the decomposition is
\begin{eqnarray}
\mathcal{T}^{00}_f&=&-\frac{i}{2}\psi^\dagger\Sigma^iD_i\psi\;\;=\;\;-\frac{1}{2}\mathcal{T}\;,\nonumber\\
\mathcal{T}^{i0}_f&=&-\frac{i}{2}\psi^\dagger\alpha^i\Sigma^jD_j\psi\;\;=\;\;-\frac{1}{2}\mathcal{T}^i\;,\nonumber\\
\mathcal{T}^{0i}_f&=&\frac{m}{2}\psi^\dagger\left(\beta\Sigma^i+\frac{i}{m}\gamma^5D^i\right)\psi\;\;=\;\;\;-\frac{m}{2}\mathcal{J}^i,\nonumber\\
\mathcal{T}^{ij}_f&=&\frac{m}{2}\psi^\dagger\alpha^i\left(\beta\Sigma^j+\frac{i}{m}\gamma^5D^j\right)\psi\;\;=\;\;
-\frac{m}{2}{\mathcal{J}}^{ij}\;,\label{dec0}
\end{eqnarray}
where fermionic field equations \eqref{Eqf} have been used once again. In \eqref{dec0} we have employed the standard notation of $\gamma$-matrices: $\beta=\gamma^0$, $\alpha^i=\gamma^0\gamma^i$ and $\Sigma^i=\gamma^5\gamma^0\gamma^i$. Thus, $\partial_\nu\mathcal{T}^{\nu\mu}_f=\partial_0\mathcal{T}^{0\mu}_f+\partial_j\mathcal{T}^{j\mu}_f$ splits into
\begin{eqnarray}
\partial_\nu\mathcal{T}^{\nu0}_f&=&-\frac{1}{2}\left[\partial_0\mathcal{T}+\partial_i\mathcal{T}^i\right]\;,\nonumber\\
\partial_\nu\mathcal{T}^{\nu
i}_f&=&\frac{m}{2}\left[\partial_0\mathcal{J}^i+
\partial_j\mathcal{J}^{ji}\right]\;,
\end{eqnarray}
and, from (\ref{div1}),
\begin{eqnarray}
\frac{\partial\mathcal{T}}{\partial t}+\overrightarrow{\nabla}\cdot\overrightarrow{\mathcal{T}}&=&e\overrightarrow{S}\cdot\overrightarrow{E}\;,\nonumber\\
\frac{\partial\overrightarrow{\mathcal{J}}}{\partial t}+\overrightarrow{\nabla}\cdot\overleftrightarrow{\mathcal{J}}&=&\frac{e}{m}\left(\overrightarrow{S}\times\overrightarrow{B}
-S_0 \overrightarrow{E}\right)\;.\label{equiv0}
\end{eqnarray}
in complete accordance with \cite{Vernes:2007}. It is worth mention that the extra two equations discussed in \cite{Vernes:2007} for the auxiliary density defined by the authors are simply the on-shell chiral equation discussed in Sect.2 and its dual version. The second of (\ref{equiv0}) is then directly identified with (\ref{paulidirac}). Thus, $\mathcal{T}^i$ and $\overleftrightarrow{\mathcal{J}}^{ij}$ are identified with the relativistic generalization of the spin density and spin current, respectively. This identification were formally shown in \cite{Vernes:2007}. Also in \cite{Vernes:2007}, the first and second order relativistic corrections to (\ref{paulidirac}) were computed and associated with a contribution to the spin-Hall effect. 

The bosonic current decomposition is much simpler since
\begin{equation}
T^{\mu\nu}_b=T\eta^{\mu\nu}\;,\label{dec1}
\end{equation}
where $T=\overrightarrow{E}\cdot\overrightarrow{B}$. Thus, equation \eqref{div2} decomposes as
\begin{eqnarray}
\frac{\partial T}{\partial t}&=&-\frac{1}{6}\overrightarrow{j}_f\cdot\overrightarrow{B}\;,\nonumber\\
\overrightarrow{\nabla}T&=&\frac{1}{3}\left(-\rho\overrightarrow{B}+\overrightarrow{j}_f\times\overrightarrow{E}\right)\;,\label{bos1}
\end{eqnarray}
where $\rho=j_f^0$.

In a great number of systems the magnetic and electric fields are mutually orthogonal implying that in general $T_b^{\mu\nu}=0$. In that case we have then
\begin{eqnarray}
\overrightarrow{j}_f\cdot\overrightarrow{B}&=&0\;,\nonumber\\
\overrightarrow{B}&=&\frac{1}{\rho}\overrightarrow{j}_f\times\overrightarrow{E}
\end{eqnarray}
Those are consistent classical relations, in fact, if $j=\rho v$, we achieve $\overrightarrow{B}=\overrightarrow{v}\times\overrightarrow{E}$, which is the electric-magnetic field relation encountered for a moving particle with constant velocity. However, it could be the case that the electric and magnetic field are not mutually orthogonal and then generate a spin flow for the photon.

\section{Conclusions}

By considering the usual gauge theory for electrodynamics we were able to provide a more fundamental physical meaning for the currents encountered in \cite{Vernes:2007}. Further, we have generalized the results of \cite{Vernes:2007} to include a dynamical electromagnetic field. Our results are described in what follows.

Instead of considering a relativistic quantum mechanical approach, we worked in a gauge theory framework by considering the usual electrodynamics action as a $U(1)$ gauge theory, both techniques being expected to be equivalent. Thus, spin currents might be important also outside spintronics. Formally, our results are based on the symmetries of QED. It is valid to remark that the gauge symmetry is of great importance in obtaining (\ref{div1}), \emph{i.e.}, the role of gauge symmetry lies in the generation of a non-conservative law of nature.

The equation that describes the relativistic generalization of spin current dynamics encountered in \cite{Vernes:2007} is given here in a Lorentz covariant form, (\ref{div1}). Moreover, this equation and also the current $\mathcal{T}^{\mu\nu}_f$, are gauge invariant, in contrast to the continuity equation for the Bargmann-Wigner tensor \eqref{barg}. This property ensures the interpretation of spin currents as physical observables of the theory. Another interesting point that emerges from the covariant description here presented is that the spin current and spin-torque transfer are components of the more general gauge invariant Bargmann-Wigner current. Moreover, in the \emph{rhs} of \eqref{div1} it is evident that the chiral current plays a fundamental role in the spin dynamics. Furthermore, it is not difficult to show that for massless fermions equation (\ref{div1}) is also valid off-shell, a property that is always welcome for a quantum description of a field theory.

An additional result is provided by the equations for the spin current $T^{\mu\nu}_b$ related to electromagnetic field, namely \eqref{bos1}. It can be visualized as a measure of non-perpendicularity between electric and magnetic fields.From \eqref{dec1} we can infer that $T$ is the scalar function that determines the photon spin flow. From the first of \eqref{bos1} we it is evident that, in a system were the magnetic field is not orthogonal to the charge current there will be a time varying photon spin flow. And from the second of \eqref{bos1} a gradient of spin emerges due to a kind of magnetic monopole force. This equation suggests then that, if the fields $\overrightarrow{E}$ and $\overrightarrow{B}$ and the current $\overrightarrow{j}$ are not mutually orthogonal, an effective magnetic monopole with charge $\rho_m = -\rho/3$ and $\vec{j}_m=\vec{j}_f/3$ emerges generating a spin gradient for the photon. Under this scenario $T$ can be interpreted as a potential function for this monopole force.

Finally, we wish to remark that, in despite of its interpretation, expression \eqref{div1} in the covariant form, or in the space-time decomposition form \cite{Vernes:2007}, was already known in the literature \cite{Sokolov:1986nk}.

\section*{Acknowledgements}

RFS is thankful to the Conselho Nacional de Desenvolvimento Cient\'{\i}fico e Tecnol\'ogico\footnote{RFS is a level PQ-2 researcher under the program Produtividade em Pesquisa, 304924/2009-1.} (CNPq-Brazil) and to the Pro-Reitoria de Pesquisa, P\'os-Gradua\c{c}\~ao e Inova\c{c}\~ao\footnote{Under the program Jovens Pesquisadores 2009, project 304.} of the Universidade Federal Fluminense (Proppi-UFF) for financial support. VJVO acknowledges Diego Gonz\'alez for fruitful discussions.

\end{document}